# Late Accretion of Ceres-like Asteroids and Their Implantation into the Outer Main Belt

Driss Takir[1†*], Wladimir Neumann[2,3,4], Sean N. Raymond[5], Joshua P. Emery[6], Mario Trieloff[4]

[1]Jacobs, NASA Johnson Space Center, Houston TX 77058, USA (driss.takir@nasa.gov).
[2]Institute of Geodesy and Geoinformation Science, Technische Universität Berlin, Kaiserin-Augusta-Allee 104-106, 10553 Berlin, Germany.
[3]Klaus-Tschira-Labor für Kosmochemie, Institut für Geowissenschaften, Universität Heidelberg, Im Neuenheimer Feld 234-236, 69120 Heidelberg, Germany.
[4]Institute of Planetary Research, German Aerospace Center (DLR), Rutherfordstr. 2, 12489 Berlin, Germany.
[5]Laboratoire d'Astrophysique de Bordeaux Université de Bordeaux - B18N Allee Geoffroy Saint-Hilaire, CS 50023 33615 Pessac Cedex, France.
[6]Northern Arizona University, Flagstaff, Arizona, USA.
†Visiting astronomer at the Infrared Telescope Facility under contract from the National Aeronautics and Space Administration, which is operated by the University of Hawaii, Mauna Kea, HI 96720, USA.
*Corresponding author. Email: driss.takir@nasa.gov.

**Abstract:**

Low-albedo asteroids preserve a record of the primordial solar system planetesimals and the conditions in which the solar nebula was active. However, the origin and evolution of these asteroids are not well-constrained. Here we measured visible and near-infrared (~0.5 – 4.0 μm) spectra of low-albedo asteroids in the mid-outer main belt. We show that numerous large (d > 100 km) and dark (geometric albedo < 0.09) asteroids exterior to the dwarf planet Ceres' orbit share the same spectral features, and presumably compositions, as Ceres. We also developed a thermal evolution model that demonstrates that these Ceres-like asteroids have highly-porous interiors, accreted relatively late at 1.5-3.5 Myr after the formation of calcium-aluminum-rich inclusions, and experienced maximum interior temperatures of < 900 K. Ceres-like asteroids are localized in a confined heliocentric region between ~3.0-3.4 AU but were likely implanted from more distant regions of the solar system during the giant planet's dynamical instability.

**Main Text:**

Dark asteroids, most of which are located in the mid-outer main belt (2.5 < a < 4.0 AU) (*1*), are thought to be leftovers from the formation of the planets (*2*) and remnants of the primary accretion of the first solar system planetesimals (*3*). These asteroids are genetically linked to carbonaceous chondrites (*4*) and are collectively called primitive asteroids. Primitive asteroids have been studied using various observational techniques, including ground-based (*e.g., 5, 6*) and space-based telescopes (*7*) and spacecraft (*8*). Using the NASA Infrared Telescope Facility (IRTF) telescope, (*5*) measured spectra (~1.9-4.1 µm) of low-albedo asteroids in the mid-outer main belt. Based on the 3-µm band shape, (*5*) have identified four main 3-µm spectral groups:

1. The sharp group has a 3-µm band consistent with CI or CM carbonaceous chondrites (e.g., *9*). Most of this group's asteroids are located closer to the Sun in the 2.5 < a < 3.3 AU region.
2. A group of asteroids that are spectrally similar to Ceres, located in the ~2.6 < a < 3.2 AU region, and has a relatively narrow feature centered at 3.05 µm superimposed on a much broader absorption feature from ~2.8 to 3.7 µm.
3. A group of asteroids that are spectrally similar to asteroid (52) Europa, located in the ~3.1 < a < 3.2 AU region and exhibit a 3-µm band centered at 3.15 µm also superimposed on a much broader absorption feature from ~2.8 to 3.7 µm. No meteorite match was found for both Ceres- and Europa-like groups in the 3-µm band (*10*).
4. The rounded group is located in the ~3.4 < a < 4.0 AU region and has a 3-µm band that is possibly attributed to water ice (e.g., *11*).

Here we present new visible and near-infrared (NIR: ~0.5-4.0 µm) reflectance spectra of 10 Ceres- and Europa-like asteroids (including new spectra of asteroid Europa) measured at IRTF. We place the astronomical observations of these large dark asteroids in the context of the thermal evolution and dynamical models and provide new interpretations for the origin and evolution of these asteroids.

**Orbital distribution of Ceres-like asteroids**

In this work, we are updating the classification of Takir and Emery (*5*) and combining Ceres- and Europa-like groups together in one group, Ceres-like group. We identified nine additional asteroids (in addition to asteroid Europa) that are spectrally similar to Ceres, have diameters greater than ~100 km with different spectral classes including C, B, P, F, and D, and geometric albedos of less than 0.09 (Supplementary Table 1). Spectra of Ceres-like asteroids exhibit a band centered around 3.05-3.15 µm (Supplementary Table 2, Column 2) mainly superimposed on a much broader absorption feature from ~2.8 to 3.7 µm and a broad feature centered ~1-1.5 µm that is generally concave-up shape at shorter wavelengths (~0.6 –2.0 µm) (Supplementary Fig. 1). The absorption band intensities at 3 µm for these asteroids range from ~2% to 8% (Supplementary Table 2, Column 3). The 3-µm band areas range from 0.012 to 0.015 µm$^{-1}$ (Supplementary Table 2, Column 4). Previously published visible spectra (0.4-0.93-µm, https://sbnapps.psi.edu/ferret/) are also included in this study to provide a deeper understanding of the surface composition of these asteroids. Supplementary Fig. 2 shows processed spectra of all newly identified large dark asteroids that are found to be consistent with those of the dwarf planet Ceres.

Fig. 1 illustrates that all identified Ceres-like asteroids are concentrated between ~3.0 and 3.4 AU. Two asteroids, (24) Themis and (65) Cybele, which were classified by Takir and Emery (*5*) in the rounded group, are reclassified in this work in the Ceres-like group because of their spectral similarity to Ceres' spectra. Asteroid (324) Bamberga is grouped by Takir and Emery (*5*), Rivkin et al. (*12*), and Rivkin et al. (*13*) as a Ceres-like object; however, more recent work has shown that this asteroid is classified in the sharp group (*14*). Bamberga's spectral shape in the ~0.5-2.5 μm region, characterized by a steep slope with a slight upward curvature around 1.5 μm, further suggests the classification of Bamberga in the sharp group. Bamberga may have a heterogeneous surface representing both groups, Ceres-like and sharp.

Except for asteroid Cybele located at 3.4 AU, all the studied asteroids are located in a narrow heliocentric region between ~3.0 and ~3.2 AU, beyond Ceres' orbit at ~2.8 AU. Fig. 1 also shows bodies located closer to the Sun whose ice melted, leading to aqueous alteration (sharp group), and those further from the Sun that contains unmelted ice (rounded group) (5).

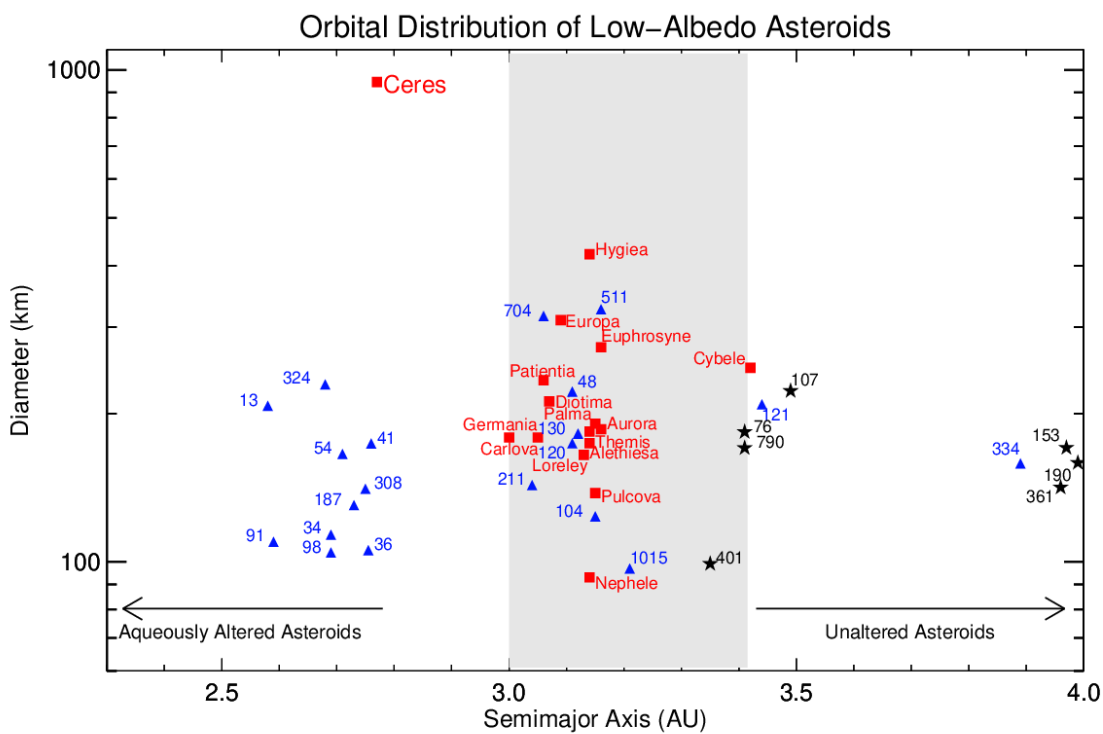

**Fig. 1. Orbital distribution of large dark asteroids.** The identified large dark asteroids (red squares) that were found to be spectrally similar to the dwarf planet Ceres are located in the ~3.0-3.4 AU heliocentric region (gray). The CM/CI-like and aqueously altered asteroids (sharp group, blue triangles) from Takir and Emery (5) are plotted at lower solar distances, and unaltered asteroids (rounded group, black stars) at greater solar distances.

**Ceres-like asteroids' thermal evolution**

Primitive asteroids are thought to be composed initially of mixtures of anhydrous materials and water ice that was later melted by heat sources such as the decay of $^{26}$Al, reacting with anhydrous materials to form $H_2O$/OH-rich minerals (15), and a carbonaceous chondritic CM- or CI-like composition. The temperature and bulk density evolution of initially water-rich small bodies were calculated using a 1D finite differences thermal evolution model (16, 17) for planetesimals heated by $^{26}$Al, $^{60}$Fe, and long-lived radionuclides. An ice-rich initial composition that leads to a material dominated by phyllosilicates after aqueous alteration (with an initial ≈13 wt.% $H_2O$ consumed completely for the aqueous alteration leading to a composition with ≈60 vol.% hydrated silicates as suggested by the water content and modal mineralogy of CM and CR chondrites) was assumed.

In particular, thermally activated compaction due to hot pressing of asteroids from an initially low density (i.e., highly unconsolidated porous structure) and a bigger size to a higher density and a smaller size was modeled to fit the bulk densities and sizes of the studied large dark asteroids. To this end, a typical initial porosity of 50% (*17*) was reduced following the change of the strain rate calculated as Voigt approximation from the strain rates of components (*17*). Material properties (thermal conductivity, density, heat capacity, etc.) corresponded to the composition assumed and were adjusted with temperature and porosity. For illustration and comparability, we associate bodies of an equal mass but different porosities (i.e., different bulk densities and sizes obtained after a model calculation) with a "reference" mass. The reference mass and accretion time intervals covered by our estimates are $10^{15}$ kg to $10^{21}$ kg (that would correspond to diameters of ~10 km to 1000 km at a grain density of 2460 kg m$^{-3}$ and zero porosity) and 1 Myr to 5 Myr relative to the formation of calcium-aluminum-rich inclusions (CAIs), respectively. Models with a low initial density (i.e., a high initial porosity) were calculated for all resulting pairs of parameters. The masses and bulk densities obtained from compaction were searched for matches with the masses and bulk densities of observed asteroids. From those matches, accretion times were derived. Fig. 2 shows the maximum temperature (a) and the average density (b) calculated as functions of reference masses and accretion time, where the mass and the accretion time $t_0$ were varied as stated above. Furthermore, the accretion times for several of the asteroids studied that result from the bulk density and mass fits are shown. See the Methods for more information about the fitting procedure of bulk densities for the studied large dark asteroids and their corresponding accretion times and uncertainties.

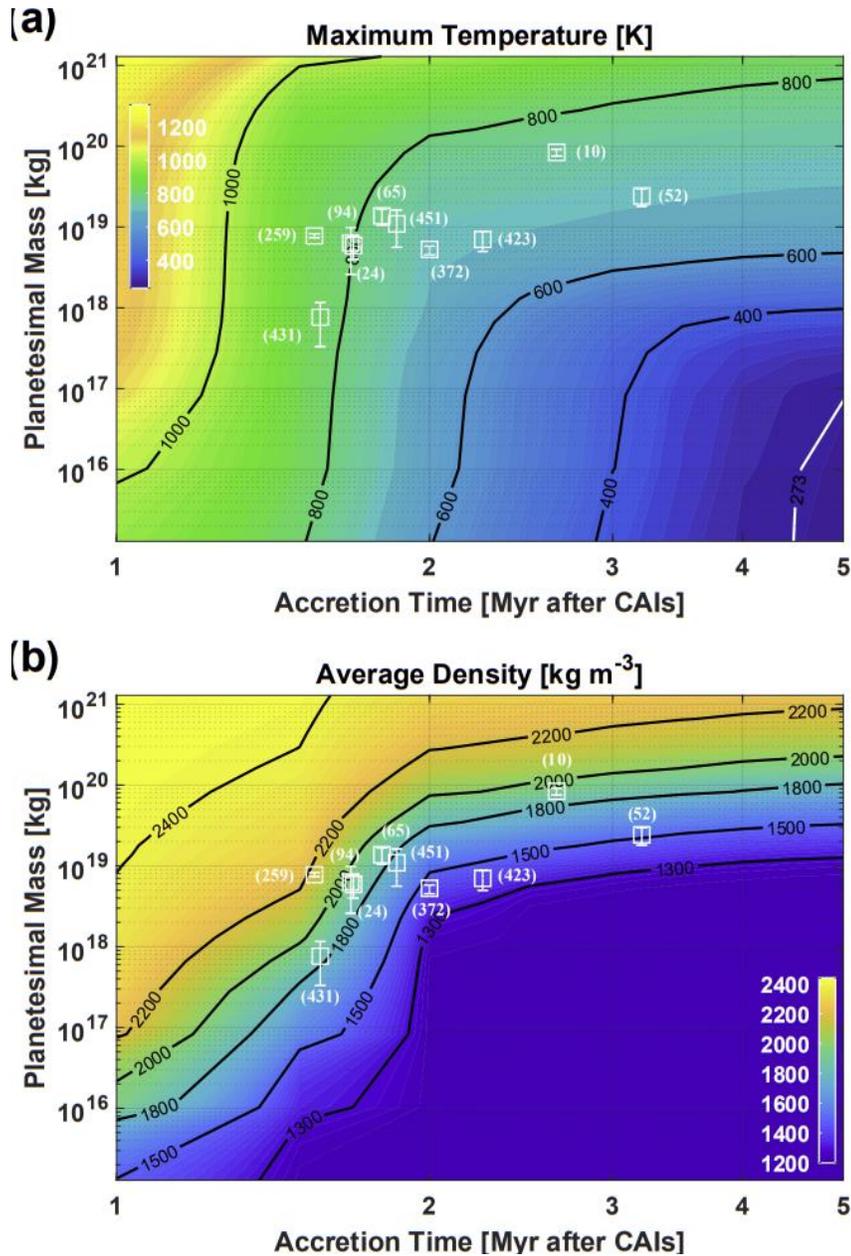

**Fig. 2. Maximum temperature (a) and density (b) for initially water-rich planetesimals as functions of the accretion time relative to CAIs and the reference mass.** Squares show fitted asteroids centered at their reference masses and accretion times. Vertical error bars show the mass uncertainty. The studied asteroids' numbers are also included in the plots. See the Methods for more details about thermal evolution modeling. Supplementary Table 3 summarizes key observed properties of asteroids fitted with thermal evolution models, their initial and intermediate properties involved in the modeling procedure, and the final calculated properties of asteroid fits and resulting accretion times derived. The error bars plotted show the mass uncertainties of the fits (see the Methods) that result from the mass estimate observation uncertainties. For the accretion time estimate uncertainty, see Supplementary Table 3.

The availability of $^{26}$Al determines the heating and compaction of planetesimals, i.e., by the accretion time $t_0$, such that maximum temperatures and structures vary enormously for $t_0 < 3$ Myr relative to CAIs. However, for a later accretion, only the mass (more precisely, the surface-to-volume ratio, i.e., the size that corresponds to this mass at a given grain density and bulk porosity) of the body determines its maximum temperature and density due to nearly constant heating by long-lived radionuclides. The models fit the asteroid bulk densities as an observational parameter and calculate bulk porosities consistent with these densities. We find that most highly porous interiors characterize the large and dark asteroids (Fig. 3) and, consequently, accreted relatively late, i.e., at 1.5-3.5 Myr relative to the formation of CAIs with a maximum temperature of < 900 K (Fig. 2). Lower bounds on the maximum temperatures can be provided by comparing masses with maximum central temperatures in Fig. 2 (a) for those five asteroids for which observational constraints (i.e., the density) do not suffice for producing accretion time fits. These lower bounds are 300, 450, 550, 620, and 650 K for (762) Pulcova, (31) Euphrosyne, (360) Carlova, (165) Loreley, and (241) Germania, respectively. The range of maximum temperatures agrees with the hydration of dry silicates and is insufficient for dehydration, in agreement with the spectra of large dark asteroids.

Contrary to what was assumed by Grimm and McSween (*15*), state-of-the-art accretion models do not produce a correlation between accretion times and heliocentric distance. The current semimajor axes should not be used to derive such objects' accretion times. However, this includes such a correlation at the original accretion location of these asteroids that might have differed considerably from the present ones according to our dynamical models.

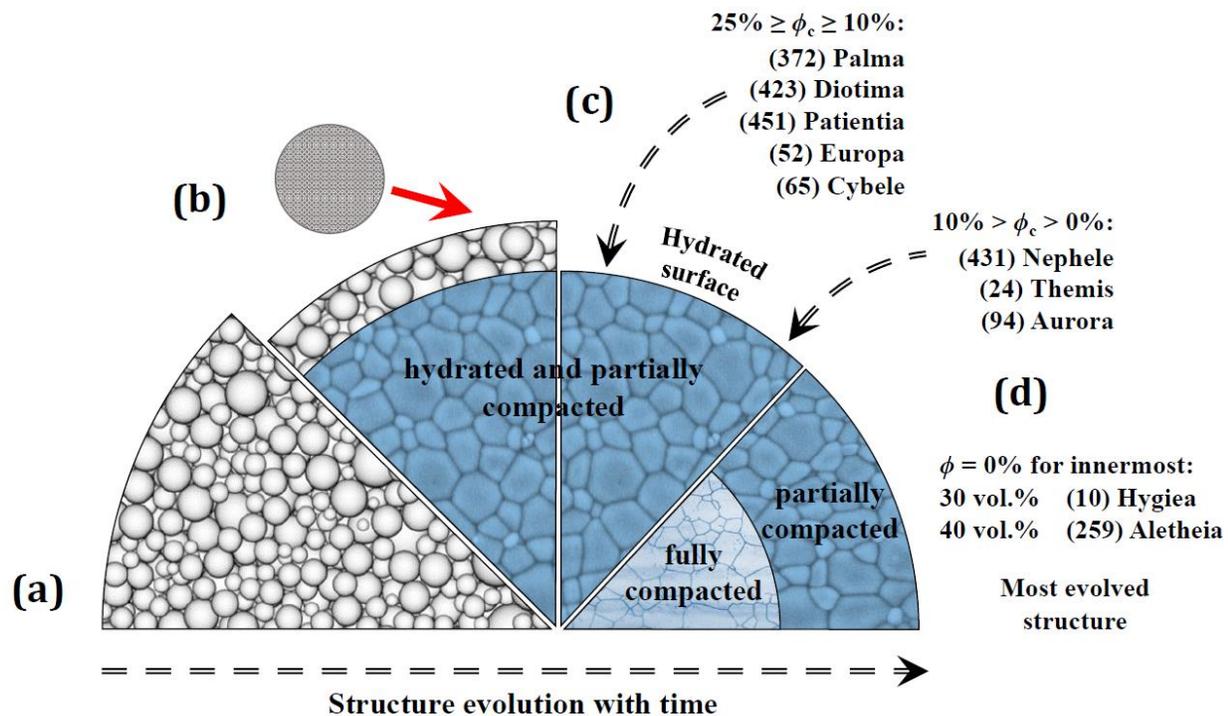

**Fig. 3. Evolution of the asteroid interior structure from left to right, as well as final structures obtained.** Asteroids form as loose agglomerates of dry dust and ice particles (a). Heating of the interior triggers hydration and progressive compaction (b). Grazing impacts then remove the remaining loose surficial layer exposing a hydrated surface (c). Some asteroids evolve to stage (d) which is characterized by fully compacted central regions. Final structures range from throughout porous ones (though partially compacted to a varying degree for different asteroids) with high central porosity of $\phi_c \approx 25\ \%$ (stage c), over strongly reduced central porosities of down to 3 % (stage c) to fully compacted interiors overlain by highly porous outer layers (stage d).

**Dynamics of Ceres-like asteroids' orbital implantation**

The planets' growth and dynamical evolution caused widespread radial mixing of small bodies (*2*). Volatile-rich asteroids were implanted from more distant orbits by being scattered inward by the giant planets and then trapped on stable orbits within the belt. This likely happened in multiple episodes (Fig. 4):

1. The giant planets' rapid gas accretion destabilized the orbits of nearby planetesimals and scattered them in all directions. A fraction was trapped in the belt under gas drag (*18*).

2. The giant planets' orbital migration acted to scatter and shepherd planetesimals. In the Grand Tack model context, the giant planets' late outward migration scattered volatile-rich planetesimals inward and implanted a fraction into the belt (*19*). Other migration pathways – such as those dominated by the inward migration of the giant planets – would also have led to implantation (*18, 20*).

3. The ice giants' accretion – from the inward migration and collisional growth of a population of ~5 Earth-mass cores (*21*) – scattered planetesimals inward, implanting a fraction in the asteroid belt due to gas drag (*22*).

4. The giant planet dynamical instability scattered the bulk of the primordial outer planetesimal disk through the inner solar system to be ejected by Jupiter (*23*). A small fraction of these very volatile-rich objects were captured onto stable orbits in the inner solar system: Jupiter's co-orbital region (*24, 25*) and the asteroid belt (*26, 27*).

These four implantation mechanisms form a rough temporal sequence: giant planet growth, migration, and instability. There is an implied gradient in the accretion ages of implanted asteroids by different mechanisms simply because planetesimals must have formed before being implanted. The feeding zones of the other mechanisms also form a rough radial progression. Jupiter and Saturn's in-situ growth (assuming no migration) implants asteroids from a zone ~5 AU wide, from ~4 to 9 AU (*18*). Accounting for the gas giants' migration and the ice giants' growth, the source region of dark asteroids would have extended from a few to ~20 AU, yet with a strong preference for objects originating between ~5 and 10 AU (*18, 20, 28*). The ice giants' growth implanted planetesimals from ~10 AU to at least 20-30 AU (*22*). Finally, the dynamical instability implanted planetesimals that accreted at ~20-40 AU (*25*).

While the exact timing of giant planet growth, migration, and instability remain uncertain, we expect later asteroid accretion times to correlate with a more distant origin and implantation by a later-occurring process. The giant planet's dynamical instability (mechanism 4) is the strongest candidate for implanting large dark asteroids. The critical piece of evidence is that these objects are concentrated in the outermost parts of the main belt, mostly between 3.0 and 3.2 AU (Fig. 1). Gas-driven capture (mechanisms 1, 2, and 3) produces a broader distribution of implanted asteroids in which it is statistically unlikely not to have captured any objects interior to Ceres' present-day orbit (*18, 19, 22*). For both mechanisms 1 and 3, the peak in the distribution of implanted Ceres-sized planetesimals is close to Ceres' actual orbit (*18, 22*). However, suppose we assume that Ceres and the large dark asteroids are part of the same population and that no large dark asteroids exist closer to the Sun than Ceres. In that case, it becomes apparent that these mechanisms cannot explain the implantation of the large dark asteroids. In contrast, implantation of asteroids during the giant planet instability (mechanism 4) is almost entirely confined to the outer belt, beyond 2.5 AU and with a steep radial number density gradient (although the exact distribution of implanted objects depends on the detailed evolution of scattered planets during the dynamical instability; *26*, *27*). This implies that the large dark asteroids originated in the 20-40 AU region (*26, 27*). The outstanding question is why the most significant object should be the innermost.

Despite their overlapping orbital distributions, dark asteroids and their more refractory-dominated counterparts (such as S-types) likely originated in disconnected regions of the solar system (*2, 29*). The carbonaceous/non-carbonaceous isotopic dichotomy measured in meteorites (*30*) suggests that their parent bodies formed concurrently but not nearby (*31*). Instead, the disk of planetesimal-forming solids was divided in two, perhaps by Jupiter's growing core, a non-planet-related pressure bump in the disk, or other processes related to the planetesimal formation (*32*). Dark primitive asteroids likely represent planetary building blocks in the outermost parts of the disk.

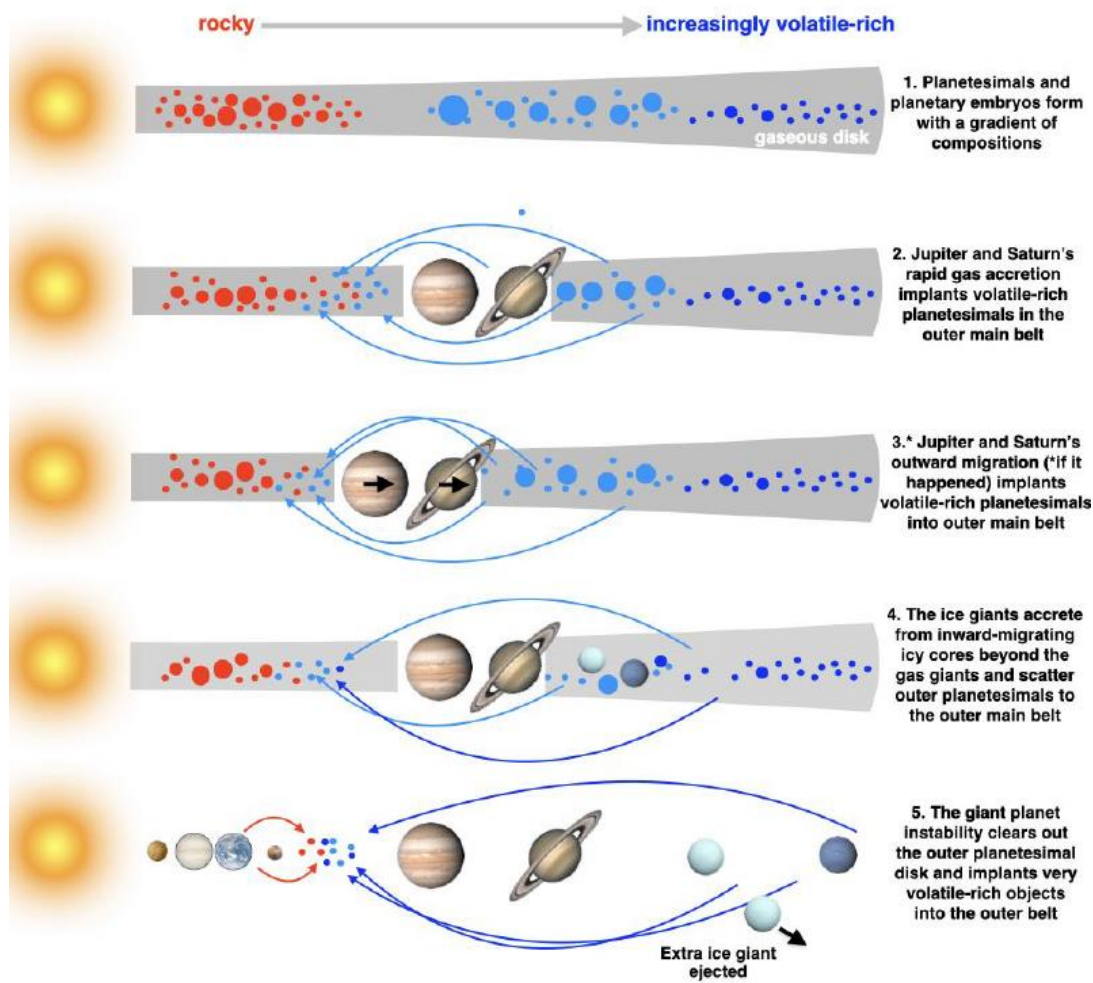

**Fig 4. Implantation of planetesimals into the asteroid belt during the planets' growth and dynamical evolution.** The disk's temperature profile established a gradient in the composition of small bodies during planetesimal formation (e.g., *15*). The presence of ammoniated phyllosilicates (*33*) indicates that Ceres (and presumably also the large dark planetesimals) likely originated beyond Saturn's orbit and perhaps much farther out. Planetesimals from the Jupiter-Saturn region were implanted during the gas giants' rapid gas accretion (*18*). The migration of the gas- and ice giants caused the source region of implanted planetesimals to expand outward (*18, 19*). The ice giants' growth implanted additional planetesimals (*22*). The giant planet's instability then implanted some asteroids from the outer planetesimal disk (*25, 26*) – this is the likely source of the large dark asteroids. The instability may have been triggered by the dispersal of the gaseous

Page **9** of **34**

disk (*34*), such that these implantation mechanisms were only separated in time by a few to ten million years. Many volatile-poor asteroids were likely implanted from the terrestrial planet-forming region (*29*).

**Surface and interior composition of large dark asteroids**
The nature of the surface compositions of Ceres, and also Ceres-like asteroids, is still under debate, and different interpretations have been put forth to explain the absorption features in these objects (*33,35,36*). Laboratory and spectroscopic experiments on meteorites representing all nine carbonaceous chondrite types found no spectral matches for these large asteroids (*37*). Previous studies of Ceres have been conducted to constrain and estimate its surface composition. Using linear mixing, (*36*) found hydroxide brucite, serpentines, and carbonates consistent with Ceres' ground-based spectra. De Sanctis et al. (*33*) estimated the surface composition of Ceres and found evidence of widespread NH3-phyllosilicates across its surface using best-fit solutions to Dawn's NIR spectra. The presence of NH3-phyllosilicates implies that material from the outer solar system was incorporated into large dark asteroids, either during their formation at a great heliocentric distance or incorporating material transported into the main belt region (*33*).

Spectra of large dark asteroids are similar to the spectra of Himalia (*38*), the largest irregular satellite of Jupiter at 5.2 AU. The irregular satellite's spectra exhibit 1- and 3-µm bands consistent with those found in Ceres-like asteroids. Additionally, NIR spectroscopic measurements of comet 67P/Churyumov-Gerasimenko by the Rosetta spacecraft (*39*) revealed that this comet's spectra exhibit a 3-µm feature that is similar to spectra of Ceres and other large asteroids. The 3-µm feature in comet 67P was attributed to ammonium salts (*39*). Mousis and Alibert (*40*) also suggested that icy ammonia, condensed in the outer solar system beyond 5 AU, could have migrated inward and incorporated into Ceres and presumably other large dark asteroids in the main belt.

Ceres has a specific composition that appears to differ from most other large dark asteroids in the main belt (*41*). In particular, Ceres has a higher density at present than most other large dark asteroids, had, potentially, a higher $H_2O$ fraction in the past that led to differentiation (*42*), and might have experienced a loss of a past water ocean (*43*). Both modeling and observations suggest a small porosity for Ceres (*41, 44*), in agreement with more vital compaction due to its larger size and more energetic thermal evolution (*45*). In contrast, current observations indicate no free water and higher porosity, and lower metamorphic temperatures for the other large dark asteroids, supported by our modeling thermal evolution section. Being much more evolved, Ceres owes its impressive geological evolution and past to its dwarf planet size and a much more extensive water action.

Is asteroid Hygiea, the fourth largest main belt asteroid, the second dwarf planet in the main belt? The Dawn mission revealed that Ceres is geologically evolved, with evidence of tectonic and cryovolcanic activities and a heavily-cratered surface (*8*). Previous observational results showed that the dwarf planet Ceres and asteroid Hygiea are spectrally similar in that they both have a 3-µm band centered at ~3.05 µm (*5*), unlike most of the studied large dark asteroids that have a 3-µm feature centered at ~3.15 µm. Additionally, Hygiea was found to have a basin-free spherical shape comparable to Ceres (*45*). The accretion time of 2.65 Myr after CAIs derived for Hygiea is earlier than the accretion times of ≥ 5 Myr after CAIs suggested for Ceres by different studies (e.g., *16* and references therein). This is consistent with less efficient cooling of Ceres through the surface than for Hygiea. If Ceres had accreted earlier, this would have led to elevated temperatures, dehydration, and, potentially, a loss of its water content. However, reducing the radiogenic energy for late accretion provides less energetic heating and compensates for a weaker cooling for Ceres. Since Ceres likely had and still has a more significant water fraction than large dark asteroids addressed here, it was explicitly not fitted. However, models with a Ceres-like water fraction would necessitate a lower grain density, shifting the density isolines in Fig. 2 (b) to more enormous reference masses and to later accretion times since the water fraction would not contain radioactive heat sources. The shapes of the density isolines in Fig. 2 (b) imply an accretion time close to or later than ~5 Myr after CAIs. Suppose Ceres and Hygeia have a similar surface composition and are of a similar origin. In that case, the accretion of planetesimals must have occurred over an extended time interval of more than 2.5 million years in the region of the solar system where both objects formed. The discovery of large dark asteroids smaller than Ceres near its orbital vicinity suggests that this dwarf planet is not unique in the main belt, and its surface composition is not driven by its vast size.

## Methods:
### Observations and data reduction

We measured asteroid spectra of large and dark asteroids with the Prism (0.7-2.52 µm) and long-wavelength cross-dispersed (LXD: 1.9-4.2 µm) modes of the SpeX (0.8 x 15 arcsec slit) spectrograph/imager at the IRTF (*46*). Tables S4 and S5 summarize the observing parameters for the studied asteroids observed between 2015 and 2020 with the prism and LXD modes of SpeX. We used the IDL (Interactive Data Language)-based spectral reduction tool Spextool (v4.0) to reduce the data (*47*).

For asteroids' LXD and Prism observations, we followed the same technique used in Takir and Emery (*5*). The spectral image frames were divided by a flat field frame measured using an internal integrating sphere. To correct for the contributions of OH line emission and the thermal emission from the sky (longward of ~2.3 µm), we subtracted spectral image frames of asteroids and the solar analog standard stars in Tables S4 and S5 (G-type stars close to asteroids on the sky at similar airmass) at beam position A from spectral image frames at beam B of the telescope. After this subtraction, the residual background was removed by subtracting the median background outside the data aperture for each channel. Spectra were extracted by summing the flux at each channel within a user-defined aperture. Asteroid spectra were divided by spectra of the solar analog measured close in airmass to remove telluric absorptions (mostly water vapor at these wavelengths). Wavelength calibration was conducted at $\lambda < 2.5$ µm using argon lines measured with the internal calibration box and at $\lambda > 2.5$ µm using telluric absorption lines.

### Thermal flux modeling and correction

The measured LXD spectra of dark asteroids show a steep increase of apparent reflectance longward of ~2.5 µm due to thermal radiation from the asteroids' surfaces. Removing the thermal excess in spectra is essential for adequately characterizing mineral absorptions in asteroids. We used the Near-Earth Asteroid Thermal Model (NEATM) (*48*) to constrain the model thermal flux longward of 2.5 µm in asteroids. NEATM is based on the Standard Thermal Model (STM) of Lebofsky et al. (*49*). The measured thermal excess was fitted with a model excess (Supplementary Figure 2) that was then subtracted from the measured thermal flux relative spectra of the asteroids (Supplementary Figure 3).

In the thermal model, we used parameters that include heliocentric and geocentric distances, visible geometric albedo, and phase angle at the time of observation obtained from the Jet Propulsion Laboratory Horizon online ephemeris generator (Supplementary Table 2). A default value for the slope parameter G of 0.15 was used for all asteroids (*50*). Because we used the V-band albedo in the thermal model, we applied a K to V scale (derived using Prism spectra) for all asteroids to reconcile the two reflectance values at the two different wavelengths. The beaming parameter was incorporated in the NEATM model to account for differences in surface roughness. To minimize a chi-squared fit to asteroids' observational data and find the surface temperature that best matches the measured thermal flux, the NEATM model iterates through several approximations of the beaming and geometric albedo. We assumed both bolometric and spectral emissivity to be 0.9 and the night sides of the asteroids to emit no thermal energy in the model.

**Calculation of the 3-µm band parameters**

The absorption feature at ~3 µm in the studied asteroids was isolated and divided by a straight-line continuum in wavelength space following the methodology used in Cloutis et al. (*51*). Two maxima determined the absorption feature's continuum at ~2.85 µm and 3.25 µm. The 3-µm band depth, $D_\lambda$, at a given wavelength, was calculated relative to the continuum using the following equation: $D_\lambda = \frac{R_c - R_\lambda}{R_c}$, (1) where $R_\lambda$ is the reflectance at a given wavelength $\lambda$, and $R_c$ is the reflectance of the continuum at the same wavelength as $R_\lambda$. Band depth uncertainty was calculated using the following equation:

$$\delta D_\lambda = D_\lambda * \sqrt{\left(\frac{\delta R_1}{R_1}\right)^2 + \left(\frac{\delta R_c}{R_c}\right)^2}, \quad (2)$$

where
$$R_1 = R_c - R_\lambda, \quad (3)$$

and

$$\delta R_1 = \sqrt{(\delta R_c)^2 + (\delta R_\lambda)^2}, \quad (4)$$

$\delta R_c$ and $\delta R_\lambda$ were derived using each wavelength's uncertainty calculated during the data reduction process.

The 3-µm band area was calculated by integrating the spectral curve below the continuum of the absorption feature. The band center was computed by applying a sixth-order polynomial fit to the central part of the absorption feature. An average of at least five measurements made by varying the positions of band maxima was used for the band area and band center calculations. Uncertainties for the band area and center were calculated by the 2σ standard deviation representing variability from the average.

**Thermal evolution model description**

The temperature and bulk density evolution of initially water-rich small bodies were calculated using a 1D finite differences thermal evolution model (*16, 17*) for planetesimals heated by $^{26}$Al, $^{60}$Fe, and long-lived radionuclides. In particular, thermally activated compaction due to the hot pressing of bodies with an initially unconsolidated porous structure was modeled to fit the bulk densities of large low-albedo asteroids under consideration and, therefore, constrain their accretion time.

**Energy balance**

A 1D finite differences thermal evolution model for planetesimals heated mainly by $^{26}$Al presented in Neumann (*16*) and Neumann (*17*) was adapted for the current study. It calculates the heating of small porous bodies, their thermal evolution, and the compaction of a mixture of dry and hydrated material from an initially unconsolidated state due to hot pressing by solving several equations that describe these processes. A non-stationary 1D heat conduction equation in spherical coordinates is discretized by the finite differences method along the spatial and temporal domain and solved for the temperature:

$$\rho c_p (1 + x_{ice} S_{ice}) \frac{\partial T}{\partial t} = \frac{1}{r^2} \frac{\partial}{\partial r}\left(kr^2 \frac{\partial T}{\partial r}\right) + Q(r,t) \quad , (1)$$

with the bulk density $\rho$, the heat capacity $c_p$, the initial fraction $x_{ice}$ and the Stefan number $S_{ice}$ for water ice, the temperature $T$, the time $t$, the radius variable $r$, and the energy source density $Q$. The energy source for the temperature change is the radioactive decay of typical short-lived radionuclides $^{26}$Al and $^{60}$Fe and long-lived $^{40}$K, $^{232}$Th, $^{235}$U, and $^{238}$U:

$$Q(r,t) = (1 - v_{ice})\rho \sum_i f_i Z_i \frac{E_i}{\tau_i} \exp\left(-\frac{t-t_0}{\tau_i}\right) \quad (2),$$

with the initial water ice volume fraction $v_{ice}$, the porosity-dependent bulk density $\rho$, the number of atoms of a stable isotope per 1 kg of the primordial material $f$, the initial ratio of radioactive and stable isotope $Z$, the decay energy $E$, the mean life $\tau = \lambda/\log(2)$, the half-life $\lambda$, and the accretion time $t_0$ of the planetesimal. See Supplementary Table 6 for the associated parameter values. A homogeneous heat source distribution within the material is assumed, while the heat source density scales further with the porosity $\phi$. The porosity is initially constant throughout the interior but develops heterogeneously with depth during the thermal evolution under the action of temperature and pressure. The temperature calculation starts from an initial value of $T_S = 230$ K throughout the planetesimal with a constant surface temperature $T_S$.

**Composition and material properties**

We assume properties similar to CM chondrites based on spectral similarities of large dark asteroids with water-rich carbonaceous chondrites. The initial composition is considered mainly a mixture of anhydrous materials and water ice that was later melted by heat sources such as the decay of $^{26}$Al, reacting with anhydrous materials to form $H_2O$/OH-rich minerals (*15*). An ice-rich initial composition that leads to a material dominated by phyllosilicates after aqueous alteration was assumed as suggested by the water content and modal mineralogy of CM chondrites (*52*), i.e., $\approx 13$ wt.% or $\approx 33$ vol.% $H_2O$ ice initially that is entirely consumed for the aqueous alteration leading to a composition with $\approx 60$ vol.% hydrated silicates and $\approx 40$ vol.% anhydrous silicates, i.e., $\approx 50$ wt.% for either. In the following, notations $x_{ice} = 0.13$, $x_{dust} = 0.87$, $v_{ice} = 0.33$, $v_{dust} = 0.67$, $x_{hyd} = 0.5$, $x_{anh} = 0.5$, $v_{hyd} = 0.6$, and $v_{anh} = 0.4$ are used. This agrees with an average grain density of $\approx 2460$ kg m$^{-3}$ calculated for a mix of water ice and olivine initially and a mix of antigorite and olivine after water consumption. This initial composition is an idealized approximation that provides material properties, such as grain density, thermal conductivity, and heat capacity, needed for the numerical procedure. This approximation does not include minor phases and trace elements. Minor phases like ammonia would influence material properties to an amount that would lead to negligible differences in the thermal evolution modeling outcomes. Therefore, they are neglected, but this neglection does not contradict the presence of ammonia in the initial composition. The grain density is at the lower end of known CM grain densities. An average grain density of 2700 kg m$^{-3}$ applied to the asteroids considered would imply current porosities of up to 50%. However, the process of cold pressing that acts during the accretion in the absence of notable heating (e.g., *53*) already reduces the porosity in planetesimals to $\approx 43\%$. Thus, we chose a grain density corresponding to no more than 43% of current porosities. The initial water content could have varied according to the range derived for CM chondrites. In general, a different water fraction would influence the energy balance in the models only via small changes in the latent heat consumption during the melting of ice. A reasonable variation in the water content would, thus, lead to negligible variations in the modeling outcome. Thus, the composition used here is reasonably representative in terms of thermal evolution and compaction behavior.

A typical initial porosity of 50% was reduced following the change of the strain rate that is calculated as Voigt approximation from the strain rates of components (*54*). While a variation of the initial porosity within a few percent appears to be reasonable, we believe that we already considered a reasonable value. Porous non-lithified material can be approximated by packings of spheres. The loosest close packing that is just stable under the application of external forces has a porosity of ~44% (*55, 56*). While planetesimals accrete from very fluffy dust aggregates with porosities of up to 90%, collisions of these aggregates lead to compaction of the material, and experiments show that such repeated micro-impacts can reduce the porosity to less than ~60% (*57, 58*) prior to the accretion of these aggregates to a planetesimal. The value of 50% used, is smaller than 60% and larger than that of the random loose packing and is, further, typically used for planetesimal thermal evolution models (e.g., (*53*) and follow-up papers).

Material properties (thermal conductivity, density, heat capacity, etc.) corresponded to the composition assumed and were adjusted with temperature, porosity, and composition. The melting of water ice is considered in a temperature interval of two degrees between $T_s = 272$ K and $T_l = 274$ K to avoid numerical issues with too sharp a phase transition at 273 K. We include the consumption of the latent heat L = $3.34 \cdot 10^5$ J K$^{-1}$ kg$^{-1}$ during the melting of the ice via a Stefan number [see Equation (1) and (*44*)] using for its calculation the heat capacity of 4200 J K$^{-1}$ kg$^{-1}$ of liquid water. The aqueous alteration and, therefore, the formation of hydrated silicates is assumed to occur quasi-instantaneously after reaching $T_l$.

The local bulk density $\rho$ is derived from the grain density $\rho_g$ by scaling it with the local volume filling factor (1-$\phi$): $\rho = (1 - \phi) \rho_g$. The equations for the heat capacity $c_p$ and the thermal conductivity $k$ are averages of the water ice and anhydrous silicates (T < $T_l$) or hydrated and anhydrous silicates (T ≥ $T_l$) contributions. The thermal conductivity k is calculated as a volume-fraction weighted geometric mean of the thermal conductivities of water ice and anhydrous silicate dust:

$$k(\phi, T) = \left(\frac{567}{T}\right)^{v_{ice}} 4.3^{v_{dust}} \left[\exp\left(-\frac{4}{0.08}\phi\right) - \exp\left(-4.4 - \frac{4}{0.17}\phi\right)\right]^{1/4} \quad (3)$$

for T < $T_l$, or of hydrated and anhydrous silicates

$$k(\phi, T) = \left(\frac{1}{0.404 + 0.000246T}\right)^{v_{hyd}} 4.3^{v_{anh}} \left[(max(1 - 2.216\phi, 0))^4 - \exp\left(-4.8 - \frac{4}{0.167}\phi\right)\right]^{1/4} \quad (4)$$

for T ≥ $T_l$.

Similarly, the heat capacity is calculated as a mass fraction-weighted arithmetic mean

$$c_p(T) = x_{ice}(185 + 7.037T) + x_{dust}(800 + 0.25T - 1.5 \cdot 10^7 T^{-2}) \quad (5)$$

for T < $T_l$, and

$$c_p(T) = x_{hyd}(0.9 - 6.3T^{-0.5} - 14600T^{-2} + 1.91 \cdot 10^6 T^{-3}) + x_{anh}(800 + 0.25T - 1.5 \cdot 10^7 T^{-2}) \quad (6)$$

for T ≥ $T_l$, where the olivine heat capacity is approximated with an H-chondritic $c_p$. For the thermal conductivities and heat capacities of the species, see (*16*) and references therein.

**Porosity**

The evolution of the volume fraction of bulk pore space, i.e., porosity $\phi$ is calculated for T ≥ $T_l$ by considering the creep of dry olivine and wet olivine as an approximation of the bulk material. It is described by non-stationary differential equations for a strain rate-stress relation.

The bulk strain rate is calculated as a volume fraction-averaged expression in the Voigt approximation (*54*):

$$\frac{\partial log(1-\phi)}{\partial t} = \dot{\varepsilon} = v_{hyd}\dot{\varepsilon}_1 + v_{anh}\dot{\varepsilon}_2 \quad (7),$$

with a diffusion creep equation for the deformation of wet olivine derived by Mei et (*59*):

$$\dot{\varepsilon}_2 = 1.2 \cdot 10^{-26}\sigma^{1.1}b^{-3}exp\left(-\frac{295000 + P \cdot 20 \cdot 10^{-6}}{8.314T}\right) \quad (8),$$

and a diffusion creep equation for the deformation of dry olivine derived by Schwenn (*60*):

$$\dot{\varepsilon}_2 = 1.26 \cdot 10^{-18}\sigma^{1.5}b^{-3}exp\left(-\frac{355640}{8.314T}\right) \quad (9),$$

with the effective stress $\sigma$, the grain size b = $10^{-5}$ m, the lithostatic pressure P, and the temperature T in K. An initial porosity of $\phi_0$ = 0.5 used is a typical value based on the porosities of the random loose and random close packings (*53, 59*). The effective stress $\sigma$ is calculated as in Neumann (*17*) for the simple cubic packing of equally sized spheres (*60*) that has a porosity of ≈ 50 %. With $\phi_0$ = 0.5 and $\rho_g$ = 2460 kg m$^{-3}$, the initial density is 1230 kg m$^{-3}$.

**Radius and resolution**

The radius $R(t)$ of the object considered changes with the bulk porosity $\phi_{bulk}(t)$ at the time $t$, obtained by integrating the local porosity over the radius $r$, according to

$$R(t) = (1 - \phi_{bulk})^{-1/3}R \quad (10),$$

where $R$ is the reference radius, i.e., the radius at zero porosity. The reference radius is used for the analysis of the results since it is representative for sets of bodies with equal mass and grain density but different porosity. All equations involved are solved on the spatial radius domain ranging from the center of the planetesimal up to its surface. The spatial grid is transformed from $0 \leq r \leq R$, with the distance from the center $r$ in m, to $0 \leq \eta \leq 1$ using the transformation $\eta := r/R(t)$. The time and space derivatives are also transformed and the transformed expressions are applied to all equations involved, such that features like Lagrangian transport of porosity and other quantities are accounted for. While the positions of the grid points between 0 and 1 are fixed, the variable values at the grid points are updated at every timestep according to the above transformations. Non-stationary equations are also discretized with respect to the time variable t and solved using the implicit finite difference method.

**Fitting Procedure**

We used asteroid bulk densities and sizes from Carry et al. (*61*) for most asteroids, Vernazza et al. (*46*) for Hygiea, and Fienga et al. (*62*) for Nephele. The total porosities of the asteroids were calculated from the densities and sizes. Note that Carry et al. (*61*) also presented porosities. However, those were macroporosities assumed to be representative of all asteroids presented in Carry et al. (*61*) for which they assumed rubble pile structure. However, based on the sizes of known rubble piles of <10 km in size (*63*), on collisional lifetimes of ≥10 km objects surpassing the age of the solar system (*64*), and on the sizes of the asteroids we consider in the present study being > 90 km, all these asteroids are likely not rubble piles and should have nearly zero macroporosity. A new macroporosity result of ~16% for the sub-km rubble pile Ryugu indicates, further, that even for tiny rubble piles, macroporosities may have been systematically overestimated in the past (*65*). Therefore, we consider that the porosity is microporosity by its nature and calculate it from the asteroid bulk densities and the meteorite analogue grain density:

$$\phi = \phi_{micro} = 1 - \frac{\rho_{asteroid,bulk}}{\rho_{meteorite,grain}} \quad (11),$$

where $\rho_{meteorite,\ grain}$ is the grain density of the meteorite analogue. As a meteorite analogue we used CM chondrites, as suggested by Carry et al. (*61*) and Clark et al. (*66*), and very similar spectral properties that indicate a similar composition. Takir and Emery (*5*), Takir et al. (*10*), and Rivkin et al. (*12*) found that the 3-μm reflectance spectra of Ceres and the largest (d > 200 km) dark asteroids have similar spectral properties and presumably compositions. We extend this statement to the asteroids investigated in the current study. Therefore, it is reasonable to assume a similar composition for the modeling. A CV analogue suggested for Aletheia and Palma and a Mesosiderite analogue suggested for Cybele by Carry et al. (*61*) would contradict asteroids' hydrated surfaces actually observed, since CV chondrites experienced very little to none aqueous alteration and have less than ~4 vol.% phyllosilicates (*52, 67, 68*), and mesosiderites are similar to igneous eucrites and diogenites in their silicate composition.

    For deriving the accretion times of the asteroids, we use an approach similar to that in Neumann (*17*), where the bulk porosity calculated for planetesimals was compared to the porosity of Ryugu's surface boulders. The goal of the fitting procedure is to find accretion time $t_0$ values that result in bulk density and diameter values that match the bulk densities and diameters of the large primitive asteroids considered. For illustration and comparability, we associate bodies of an equal mass but different porosities (i.e., also different bulk densities and sizes obtained after a model calculation) with a "reference" diameter, i.e., a theoretical diameter at zero porosity. Note that of illustration and comparability, we associated such bodies with a reference mass in the main paper since the reference diameter-scale is effectively equivalent to a mass-scale at a uniform grain density. The reference diameter and accretion time intervals covered by our calculations are 10 km to 1000 km and 1 Myr to 5 Myr relative to the formation of calcium-aluminum-rich inclusions (CAIs), respectively. Unlike previous thermal modeling investigations that assumed a correlation of the accretion times with heliocentric distance (*15*), this work showed that the accretion times of large dark asteroids derived are not found to correlate in any way with their current semimajor axes and that the current semimajor axes should not be used for deriving accretion times of such objects. Therefore, the accretion time is a free parameter, its earliest value being limited by the formation of CAIs and the latest by the end time of planetesimal accretion. Thermal evolution and compaction models with a low initial density (i.e., a high initial porosity) and bigger initial diameters than reference diameters (as prescribed by the initial porosity) were calculated for all resulting pairs of parameters.

    Within the compaction procedure, a body of a given mass and porosity (i.e., bulk density) becomes less porous (i.e., denser) due to compaction via creep. The mass remains constant, the diameter becomes smaller, the porosity becomes smaller, the bulk density becomes higher, and all values are calculated consistently. Due to the boundary condition of $T_S = 230$ K for the planetesimal surface, a relatively thin superficial regolith layer is retained, where water ice does not melt at any time during the evolution and where the initial composition of water ice and dry silicates does not change. This does not agree with the spectral observations of asteroids in the present study. In particular, the structure of this layer is that of an unsintered low-strength loose particle agglomerate. Therefore, such layers must have been ablated by impacts, such that hydrated and higher-strength partially sintered material was exposed. Removing thin upper layers by impacts is also reasonable since asteroid families are observed, and they likely formed in a similar way. Assuming that those impacts that occur after the cessation of the compaction remove the primordial low-strength layer that would partly evaporate and partly disperse, we consider only the interior of planetesimals where hydrated and at least partially lithified material was able to form and calculate the average density of the interior. The average density is a value that is bounded above by the grain density (approached only for nearly complete compaction) and below

by the initial density (retained if no compaction occurs at all), and its value depends on the thermal evolution experienced by a planetesimal. The sizes of the hydrated parts of planetesimals and their bulk densities were then searched for matches with the asteroids' sizes and bulk densities. Such matches already have the bulk densities and sizes of the asteroids. For a given asteroid, a match can occur only for some unique accretion time, such that an accretion time can be assigned to this asteroid. For the cases shown in Fig. 2, the primordial layer is typically a few 100-meters thick at most.

**Accretion time and reference diameter (or mass) uncertainties**

Our analysis is concentrated on the size and density values currently provided within the margin of error (*61*). While the size uncertainties are rather negligible, the density values have, in most cases, significant uncertainties of up to 60%. In some cases, one or both the lowest and the highest density values available are outside of the interval spanned by the initial density of 1230 kg m$^{-3}$ and the grain density of 2460 kg m$^{-3}$. Furthermore, since the thermal evolution of planetesimals that accrete after the extinction of $^{26}$Al (a few half-lives) and are heated only by long-lived radionuclides is very similar regardless of the accretion time, the bulk density isolines reach plateaus (see Fig. 2) where both peak temperature and bulk density isolines become nearly horizontal). Therefore, if plotted, some error bars for the accretion time would extend infinitely in time toward the present day. This is not reasonable since accretion times are limited by the time when the accretion of planetesimals ended. The results of our modeling inherit the shortcomings of the available data. The resulting uncertainties in the accretion times derived can be constrained only for a few cases. For (259) Aletheia, an uncertainty of +0.1/-0.6 Myr can be determined for $t_0$. In other cases, an estimate only towards either an earlier or a later formation time can be provided (or even none): (423) Diotima: -0.6 Myr; (52) Europa: -1.5 Myr; (372) Palma: -0.1 Myr; (451) Patientia: -0.9 Myr; (10) Hygiea: -0.7 Myr; (65) Cybele: -0.4 Myr relative to CAIs.

The maximally possible uncertainties for the reference diameters, thus, masses, arise from the mass estimate uncertainties. They were obtained using the equation:

$$D = 2 \left( \frac{3}{4\pi} \frac{M}{\rho_{meteorite,grain}} \right)^{1/3} \quad \text{meters} \quad (12)$$

and vary from <7 km (Aletheia) to 69 km (Patientia), corresponding to radius differences of ~3 km to ~35 km. The reference diameter spread for each asteroid is shown in Fig. 2 with vertical error bars. Our fitted reference diameters are nearly centered within these intervals, with slight tendencies towards the upper bounds.

**Data Availability:**
All data are available in the main text or the supplementary materials. Correspondence and requests for other materials should be addressed to DT (driss.takir@nasa.gov).

**Acknowledgments:**
D.T. acknowledges support by NASA's Solar System Observations grant NNX17AJ24G. J. S.N.R. thanks the CNRS' PNP and MITI programs for their support. W. N. acknowledges support by the Deutsche Forschungsgemeinschaft (DFG), project number 434933764. W. N. and M. T. acknowledge support by Klaus Tschira Foundation. We thank Benoit Carry for providing the unpublished density values of some large dark asteroids used in this study. We also wish to thank the NASA IRTF staff for their assistance with asteroid observations. Spextool software is written and maintained by Mike Cushing at the University of Toledo, Bill Vacca at SOFIA, and Adwin Boogert at NASA InfraRed Telescope Facility (IRTF), Institute for Astronomy, University of Hawai'i. NASA IRTF is operated by the University of Hawai'i under contract NNH14CK55B with NASA.

**Supplementary Material:**

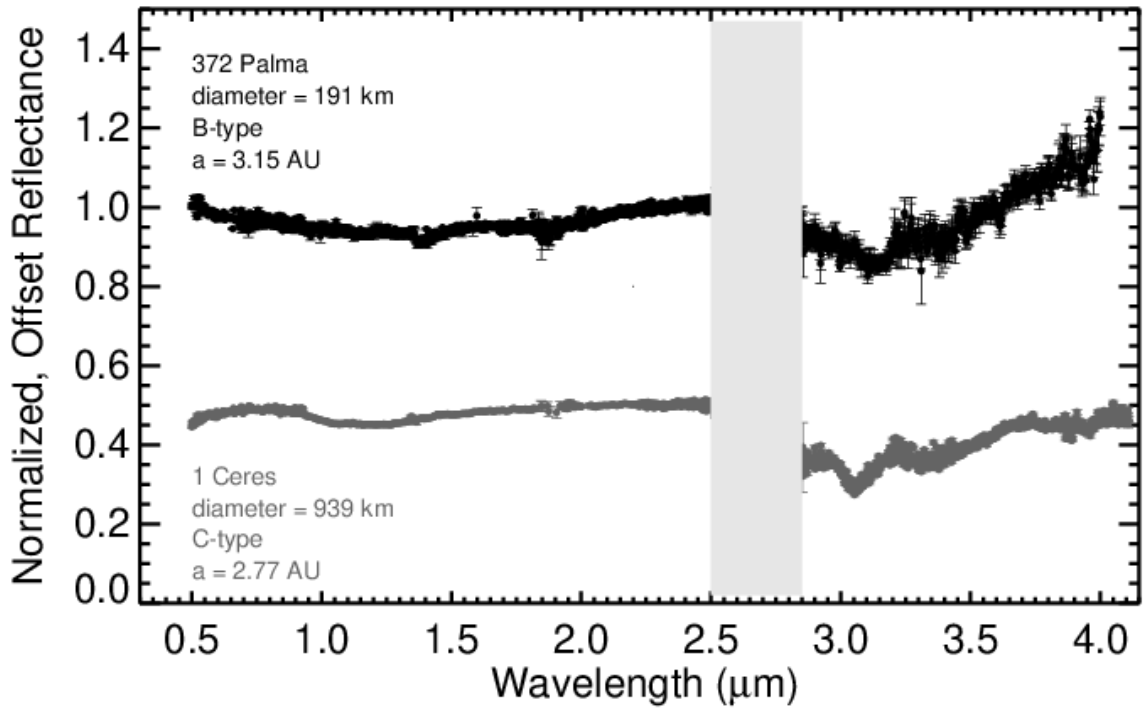

**Fig. 1. Near-infrared reflectance spectrum (binned x 6) of asteroid 372 Palma is constant with the dwarf planet 1 Ceres.** Palma's spectrum has a broad 1-µm absorption feature and an absorption band centered around 3.15 µm, consistent with Ceres' spectrum (MITHNEOS MIT-Hawaii Near-Earth Object Spectroscopic Survey, (1)). The gray bar marks wavelengths of strong absorption by water vapor in Earth's atmosphere. See Supplementary Fig. 2 for spectra of all other newly identified large dark asteroids. Uncertainties were computed by Spextool software (2) using the Robust Weighted Mean algorithm with a clipping threshold of 8 (sigma). The value at each pixel is the weighted average of the good pixels, and the propagated variance gives the uncertainty.

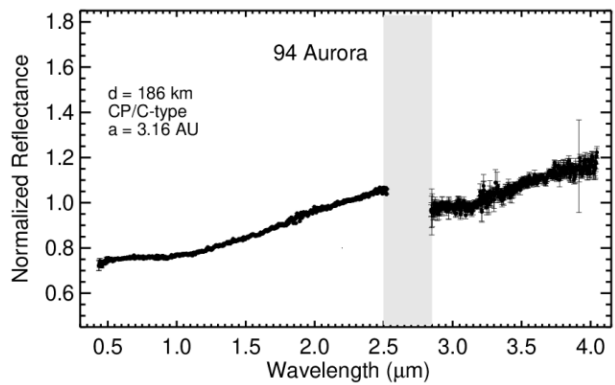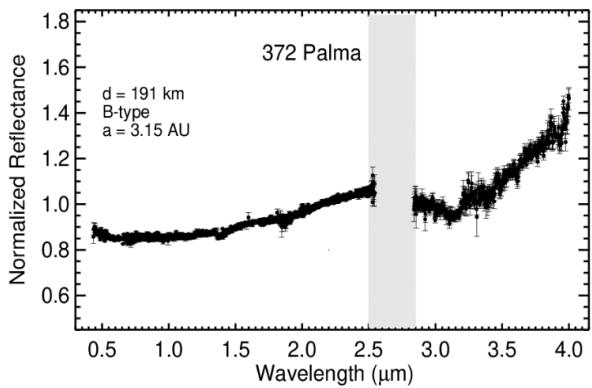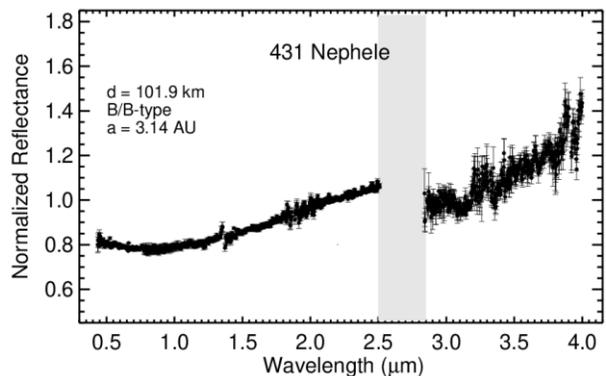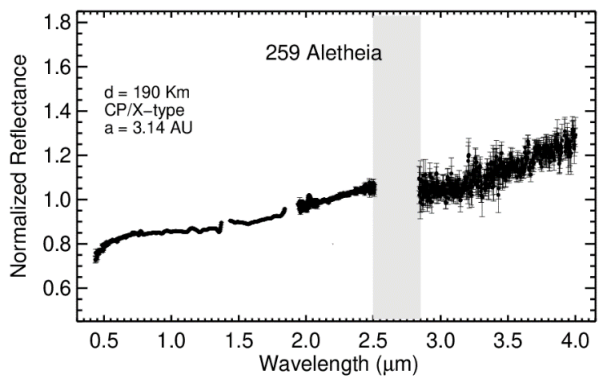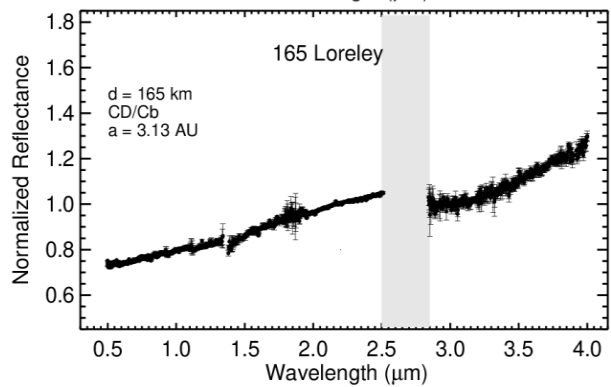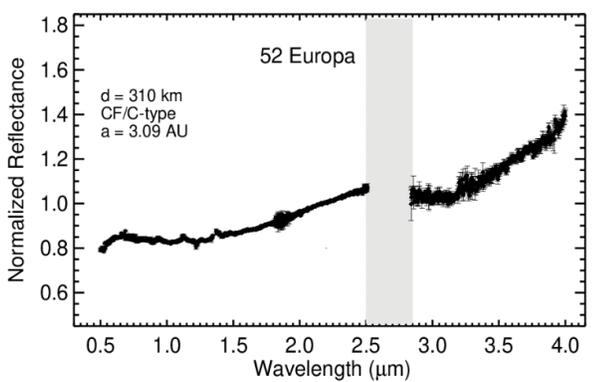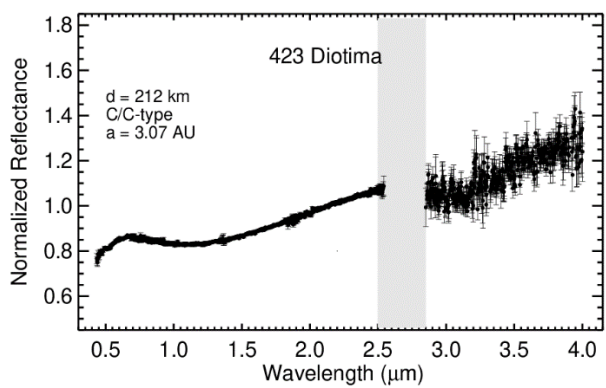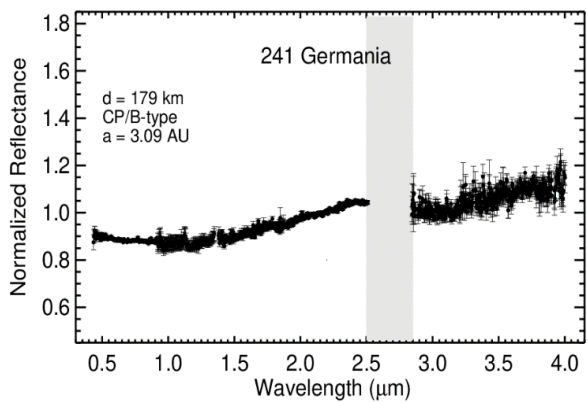

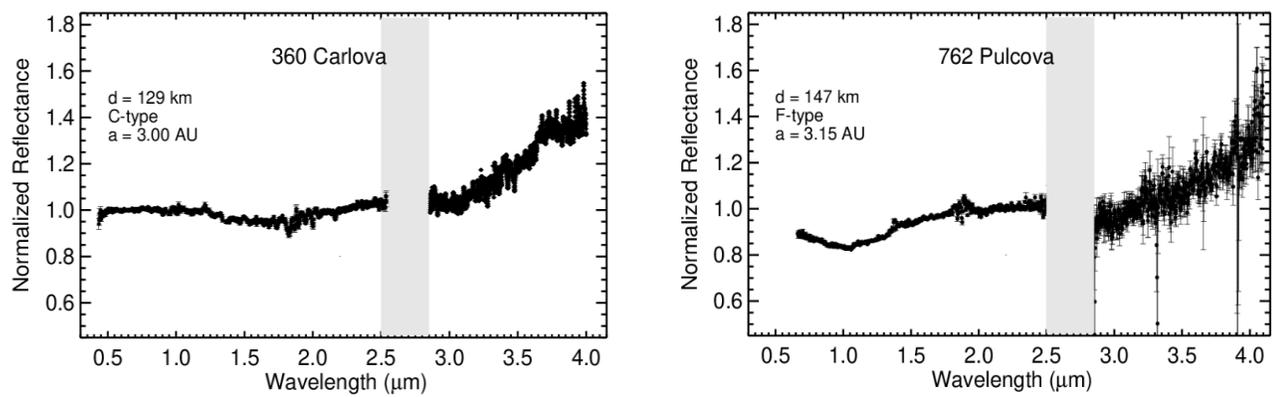

**Fig. 2. Near-infrared (binned x 6) reflectance spectra of newly identified large dark asteroids found to be consistent with the dwarf planet 1 Ceres.** All spectra have been normalized to unity at 2.2 µm. The gray bar on each plot marks wavelengths of strong absorption by water vapor in Earth's atmosphere. Asteroid (52) Europa was reobserved in this study, and its spectra were found to be similar to the ones published in Takir and Emery (3). Uncertainties were computed by Spextool software using the Robust Weighted Mean algorithm with a clipping threshold of 8 (sigma). The value at each pixel is the weighted average of the good pixels, and the propagated variance gives the uncertainty.

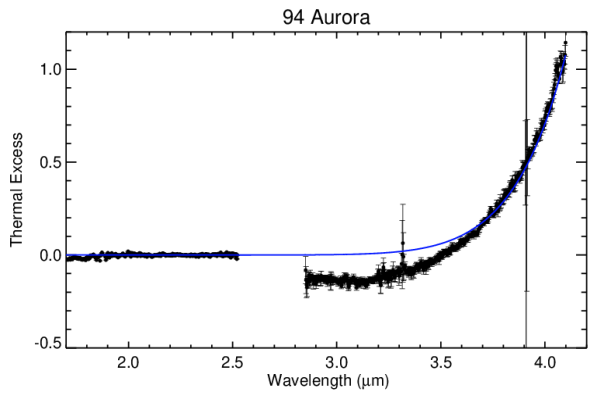
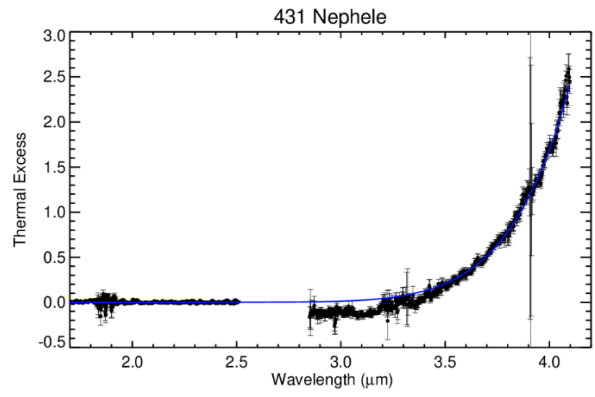
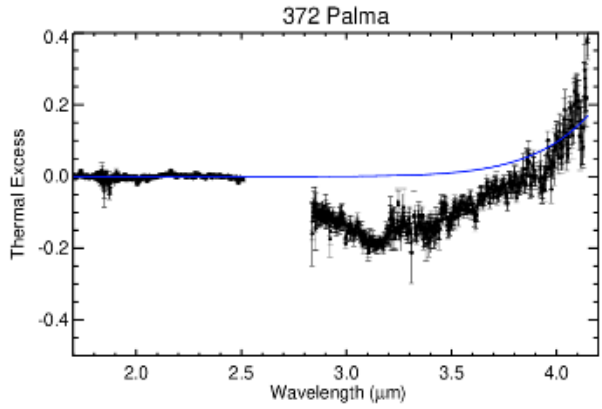
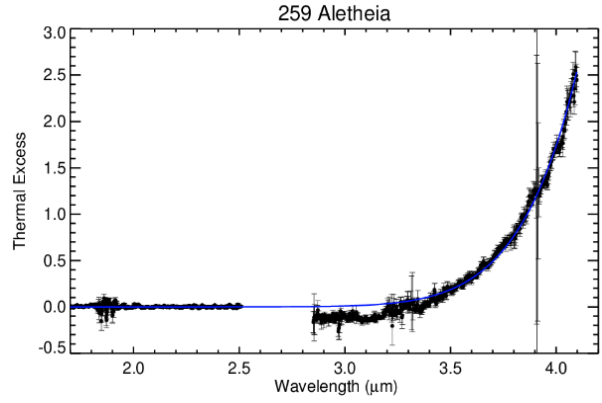
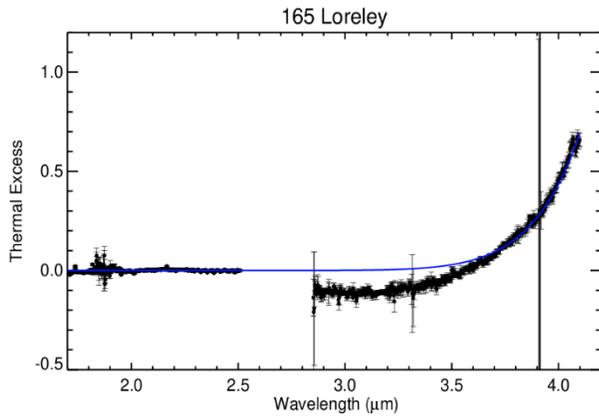
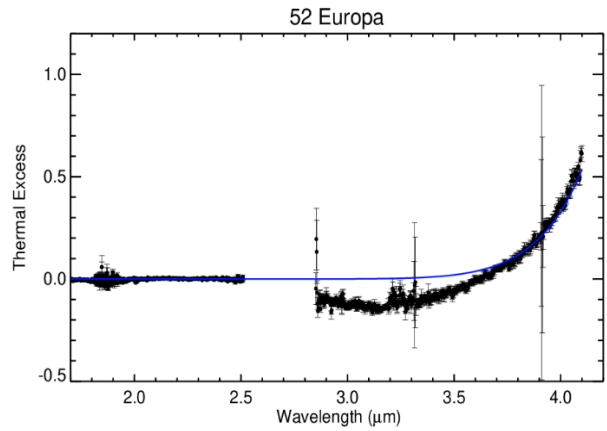
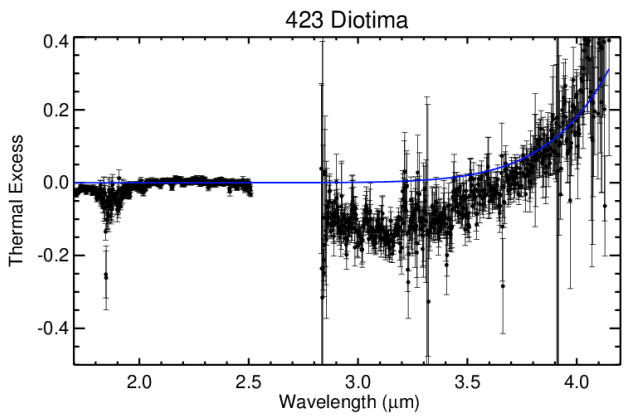
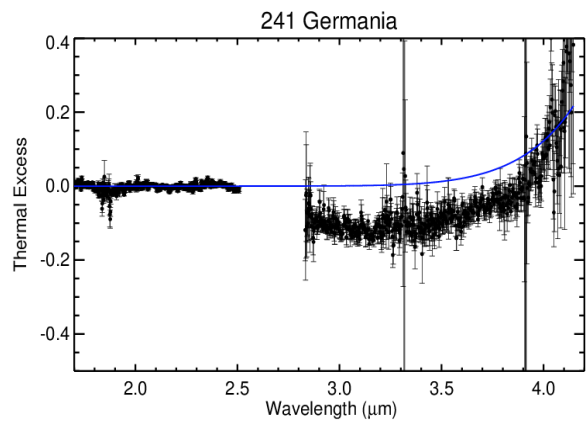

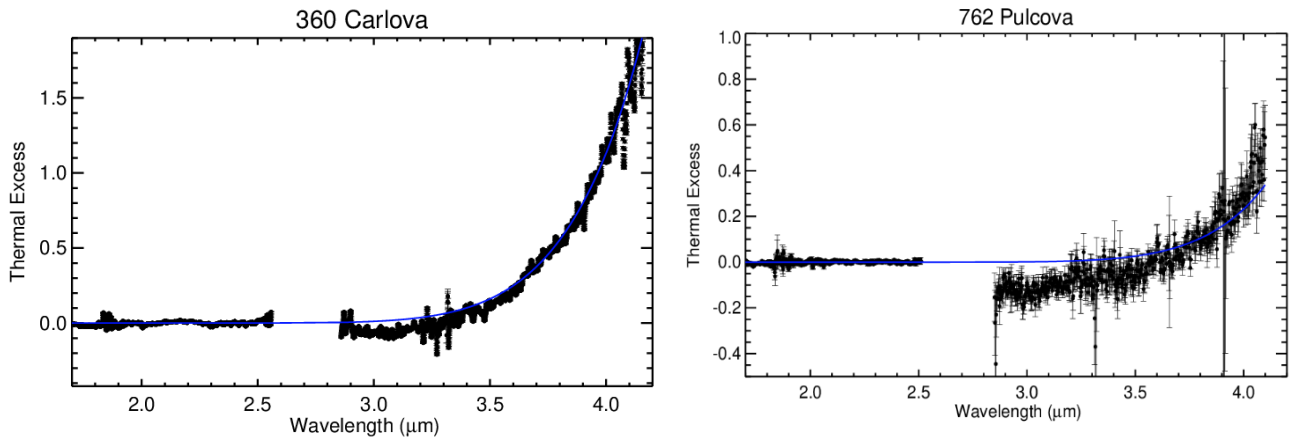

**Fig. 3. Thermal excess and NEATM's best thermal model (in blue) for the studied large dark asteroids.**
Uncertainties were computed by Spextool software using the Robust Weighted Mean algorithm with a clipping threshold of 8 (sigma). The value at each pixel is the weighted average of the good pixels, and the propagated variance gives the uncertainty.

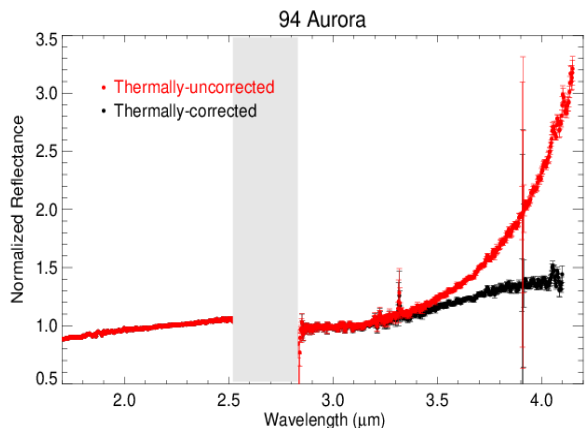
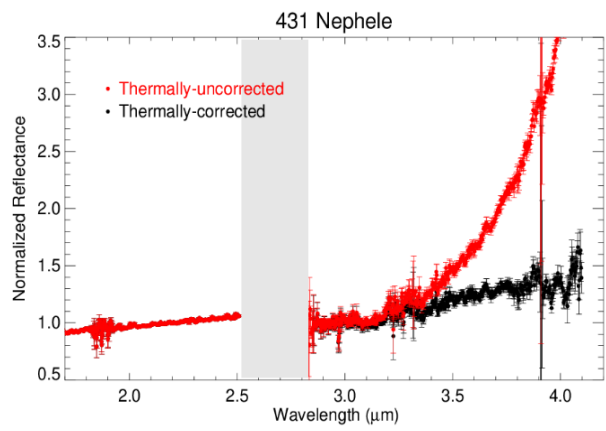
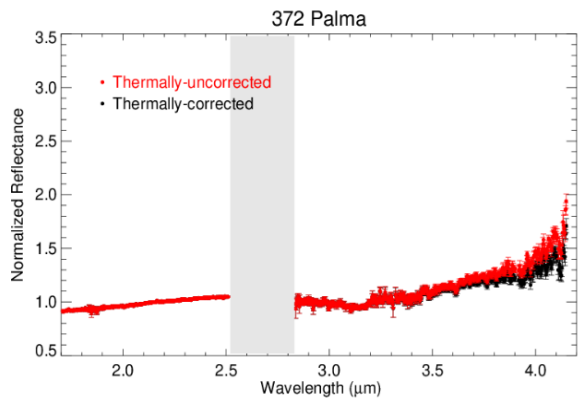
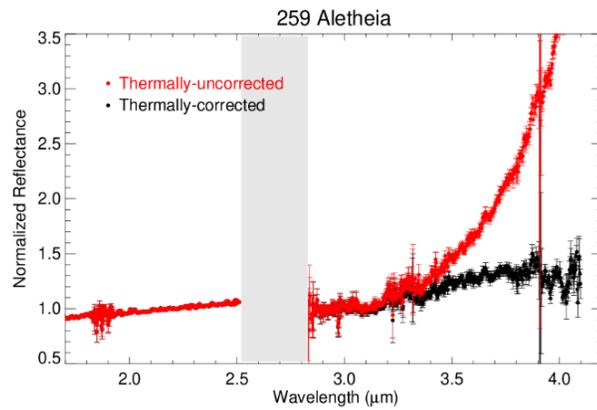
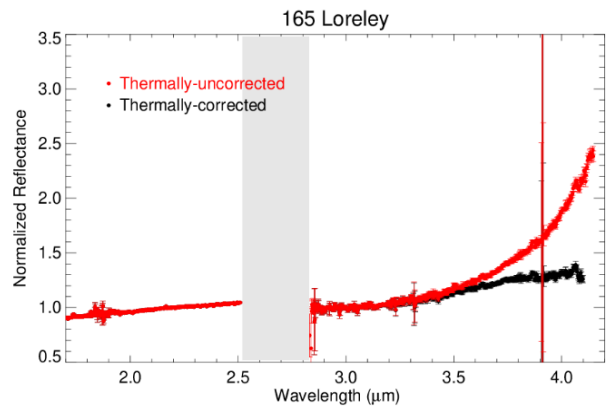
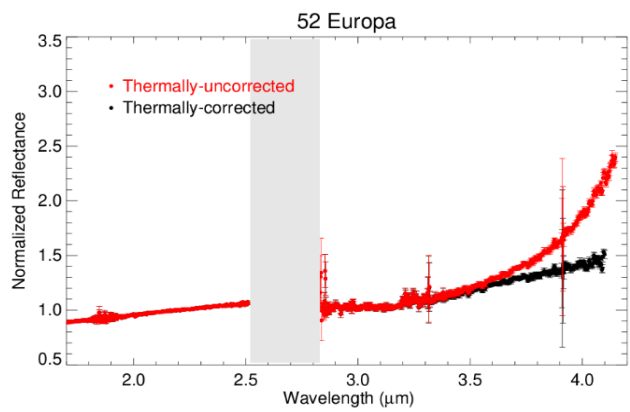
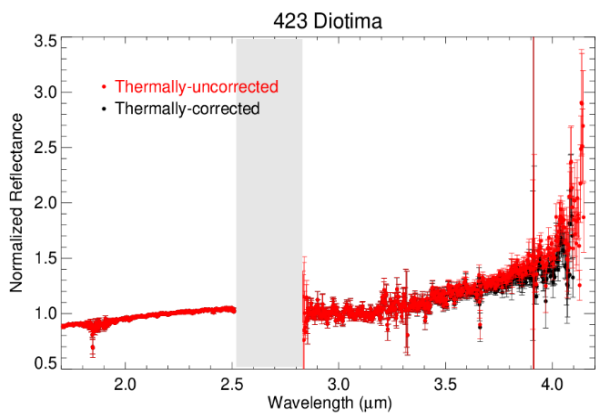
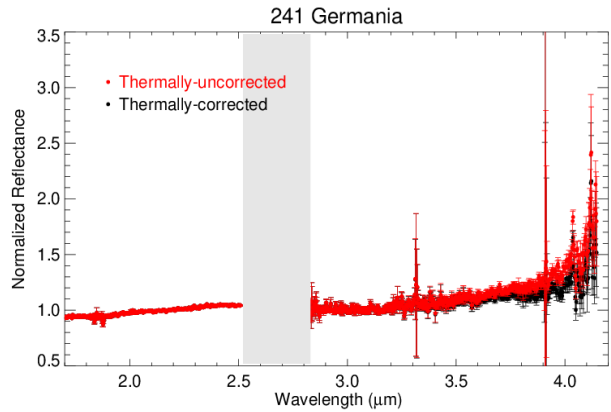

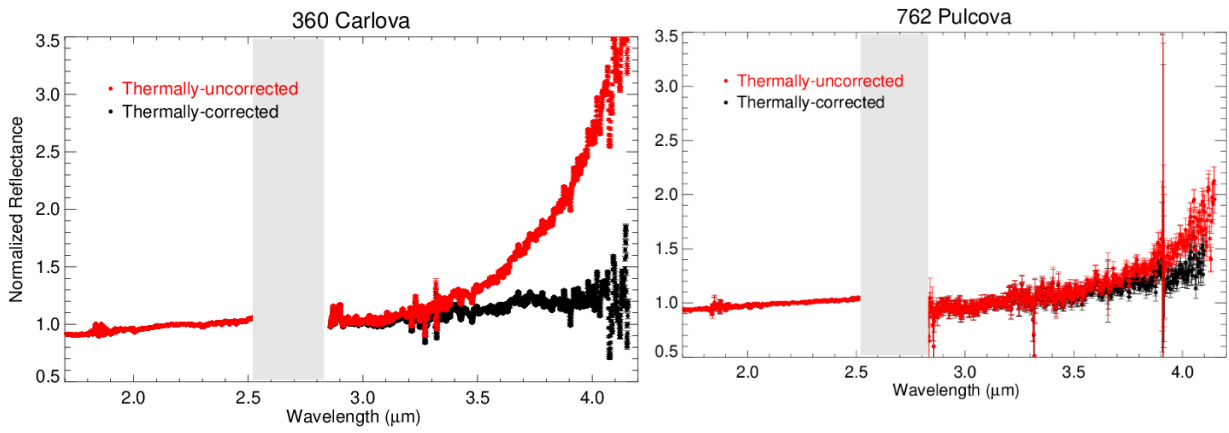

**Fig. 4. Thermally uncorrected (red) and corrected (black) spectra of large dark asteroids.** The gray bar marks wavelengths of strong absorption by water vapor in Earth's atmosphere. Uncertainties were computed by Spextool software using the Robust Weighted Mean algorithm with a clipping threshold of 8 (sigma). The value at each pixel is the weighted average of the good pixels, and the propagated variance gives the uncertainty.

**Table 1.** Physical properties of the large dark asteroids included in this study. Source: http://ssd.j.nasa.gov/sbdb.cgi. Asteroid (65) Cybele was observed by Licandro et al. (*4*); asteroid 24 Themis was observed by Rivkin and Emery (*5*) and Campins et al. (*6*); asteroids 31 Euphrosyne, 10 Hygiea, and 451 Patientia were observed by Takir and Emery (*3*).

| Asteroid | a (AU) | Eccentricity | Inclination (deg) | diameter (km)* | Spectral Class | Albedo | Average Density (kg m$^{-3}$) * |
|---|---|---|---|---|---|---|---|
| 65 Cybele | 3.42 | 0.11 | 3.57 | 248.29 ± 17.59 | P/Xc | 0.071 | 1700±520 |
| 94 Aurora | 3.16 | 0.09 | 7.97 | 186.35 ± 8.84 | CP/C | 0.040 | 1830 ± 1100 |
| 31 Euphrosyne | 3.16 | 0.22 | 26.27 | 272.82 ± 8.85 | C | 0.054 | -- |
| 372 Palma | 3.15 | 0.25 | 23.82 | 191.12 ± 2.68 | BFC/B | 0.059 | 1400 ± 180 |
| 431 Nephele | 3.14 | 0.17 | 1.83 | 93.00 ± 4.00 | B/B | 0.055 | 1767±1007 |
| 259 Aletheia | 3.14 | 0.12 | 10.81 | 190.05 ± 6.82 | CP/X | 0.043 | 2160 ± 260 |
| 10 Hygiea | 3.14 | 0.11 | 3.83 | 421.60 ± 25.69 | C | 0.072 | 1945±265 |
| 24 Themis | 3.14 | 0.12 | 0.75 | 183.84 ± 11.40 | C/B | 0.067 | 1810 ± 670 |
| 762 Pulcova | 3.15 | 0.11 | 13.10 | 138.40 ± 5.96 | F | 0.040 | 1000 ± 140 |
| 165 Loreley | 3.13 | 0.09 | 11.23 | 164.92 ± 8.14 | CD/Cb | 0.047 | -- |
| 52 Europa | 3.09 | 0.11 | 7.48 | 310.21 ± 10.34 | CF/C | 0.057 | 1500±400 |
| 423 Diotima | 3.07 | 0.04 | 11.24 | 211.64 ± 16.02 | C/C | 0.067 | 1390 ± 500 |
| 451 Patientia | 3.06 | 0.07 | 15.20 | 234.42 ± 10.17 | CU | 0.076 | 1600 ± 800 |

*The diameter and average density are from Carry (*7*) and Vernazza et al. (*8*) as well as Fienga et al. (*9*).

**Table 2.** Band depth, center, and area calculations with uncertainties for the studied large dark asteroids. Source: http://ssd.jpl.nasa.gov/horizons.cgi. This table also includes asteroid thermal model inputs used for the studied large and dark asteroids.

| Asteroid | 3-µm Band Center (µm) | 3-µm Band Depth (%) | 3-µm Band Area (µm$^{-1}$) | Rau (AU) | Dau (AU) | Phase (deg) | Abs. Mag. | Rotation Period (h) | K/V* | Temp** (K) |
|---|---|---|---|---|---|---|---|---|---|---|
| 360 Carlova | 3.05±0.06 | 5.68±0.43 | 0.012±0.003 | 2.51 | 1.59 | 9.65 | 8.56 | 6.18 | 0.99 | 253.4 |
| 241 Germania | 3.11±0.03 | 3.79 ±0.74 | 0.010±0.005 | 3.34 | 2.41 | 6.72 | 7.78 | 15.51 | 0.82 | 86.9 |
| 423 Diotima | 3.12±0.01 | 5.93 0.86 | 0.015±0.006 | 3.18 | 2.28 | 8.74 | 7.34 | 4.77 | 1.37 | 151.7 |
| 52 Europa | 3.15±0.01 | 3.89 ±0.62 | 0.008±0.004 | 2.89 | 3.22 | 18.09 | 6.48 | 5.63 | 1.22 | 228.2 |
| 762 Pulcova | 3.01±0.01 | 6.12±0.65 | 0.011±0.005 | 3.26 | 2.33 | 7.70 | 8.04 | 5.84 | 1.05 | 222.48 |
| 165 Loreley | 3.07±0.04 | 1.94 ±0.47 | 0.006±0.005 | 2.90 | 1.90 | 3.15 | 7.88 | 7.23 | 1.33 | 180.5 |
| 431 Nephele | 3.13±0.01 | 7.48±0.57 | 0.014±0.003 | 3.14 | 2.60 | 1.60 | 8.94 | 13.53 | 1.27 | 228.1 |
| 259 Aletheia | 3.13±0.01 | 3.98±0.26 | 0.009±0.002 | 3.52 | 2.55 | 3.20 | 7.90 | 8.14 | 1.23 | 215.5 |
| 372 Palma | 3.13±0.01 | 7.48±0.57 | 0.014±0.003 | 3.22 | 2.22 | 2.02 | 7.46 | 8.57 | 1.16 | 173.1 |
| 94 Aurora | 3.13±0.01 | 3.43±0.37 | 0.006±0.002 | 2.97 | 2.99 | 19.19 | 7.66 | 7.22 | 1.35 | 197.3 |

* K-band to V-band scale, applied to spectra to reconcile the two reflectance values at two different wavelengths. Rau is the heliocentric range, and Dau is the geocentric range.
** Temperatures corresponding to optimum thermal models.

**Table 3.** Key properties of asteroids fitted with thermal evolution models (columns 2-7), initial diameters (column 8), intermediate (columns 9-11), and final (columns 13-17) properties of asteroid fits, as well as the accretion times derived. Intermediate objects are slightly larger than the final objects since they have a non-hydrated and non-lithified surface layer of less than a few 100 m. After losing this thin layer by impacts, slightly smaller objects that reproduce the observed properties of asteroids are obtained. The current porosities were calculated from the masses and volumes (*7, 8, 9*) assuming a CM-like grain density of 2460 kg m$^{-3}$ for all asteroids based on their spectral similarity and meteorite analogues provided in Carry et al. (*7*). Note that Carry et al. (*7*) assumed rubble pile structures and neglected the microporosity, while we provide the total porosity in the table. The theoretical reference diameters are diameters at zero porosity. The initial bulk density is 1230 kg m$^{-3}$, and the initial porosity is 50% for all objects. The accretion times derived include uncertainties for those cases for which they could be estimated.

| Asteroid | Current semi-major axis | Current bulk density | Current porosity | Current diameter | Ref. diameter | Ref. mass | Fit initial diameter | Fit intermediate density | Fit intermediate porosity | Fit intermediate diameter | Fit final density | Fit final porosity | Fit final diameter | Fit ref. diameter | Fit ref. mass | Derived accretion time |
|---|---|---|---|---|---|---|---|---|---|---|---|---|---|---|---|---|
| | AU | kg m$^{-3}$ | % | km | km | kg | km | kg m$^{-3}$ | % | km | kg m$^{-3}$ | % | km | km | kg | Ma |
| 10 Hygiea | 3.14 | 1945±265 | 21 | 434 | 401.3 | 8,31E+19 | 506.5 | 1941 | 21.1 | 435.0 | 1946 | 20.9 | 434 | 402 | 8,37E+19 | 2.65 -0.7 |
| 24 Themis | 3.14 | 1810±670 | 26 | 183.8 | 166 | 5,89E+18 | 209 | 1795 | 27.0 | 184.5 | 1800 | 26.8 | 184 | 166 | 5,89E+18 | 1.69 |
| 52 Europa | 2.89 | 1500±400 | 38 | 310.2 | 264.4 | 2,38E+19 | 335 | 1519 | 38.3 | 312.4 | 1520 | 38.2 | 310.4 | 266 | 2,42E+19 | 3.2 -1.5 |
| 65 Cybele | 3.43 | 1700±520 | 31 | 248.3 | 219.4 | 1,36E+19 | 277 | 1706 | 30.7 | 248.6 | 1709 | 30.5 | 248 | 220 | 1,37E+19 | 1.8 -0.4 |
| 94 Aurora | 2.97 | 1830±1100 | 26 | 186.4 | 169.1 | 6,23E+18 | 214 | 1827 | 25.7 | 187.7 | 1831 | 25.6 | 187.2 | 170 | 6,33E+18 | 1.68 |
| 259 Aletheia | 3.52 | 2160±260 | 12 | 190.1 | 182.2 | 7,79E+18 | 229 | 2169 | 11.2 | 190.5 | 2176 | 11.5 | 190 | 182 | 7,77E+18 | 1.55 +0.1/-0.6 |
| 372 Palma | 3.22 | 1400±180 | 43 | 191.1 | 158.7 | 5,15E+18 | 201 | 1425 | 42.1 | 191.9 | 1427 | 42 | 191.2 | 160 | 5,28E+18 | 2 -0.1 |
| 423 Diotima | 3.18 | 1390±500 | 43 | 211.6 | 175.1 | 6,91E+18 | 222 | 1398 | 43.2 | 212.5 | 1400 | 43.1 | 211.6 | 176 | 7,02E+18 | 2.25 -0.6 |
| 431 Nephele | 3.14 | 1767±1007 | 28.2 | 93 | 83.3 | 7,45E+17 | 106 | 1771 | 28 | 93.7 | 1777 | 27.7 | 93.3 | 84 | 7,63E+17 | 1.57 |
| 451 Patientia | 3.06 | 1600±800 | 35 | 234.4 | 203.8 | 1,09E+19 | 257 | 1610 | 34.6 | 235 | 1613 | 34.4 | 234.4 | 204 | 1,09E+19 | 1.86 -0.9 |

**Table 4.** Observing parameters for asteroids observed in this study with the LXD mode of SpeX at NASA IRTF.

| Asteroid | Date (UT) | Time (UT) | Airmass | Standard star | Spectral type | B-V[*] | V-K[*] |
|---|---|---|---|---|---|---|---|
| 94 Aurora | 10/28/19 | 12:51-15:31 | 1.19-2.84 | SAO 75554 | G0 | 0.63 | 1.66 |
| 372 Palma | 08/21/16 | 11:12-13:13 | 1.13-1.44 | HD 210105 | G2V | 0.62 | 1.50 |
| 431 Nephele | 02/09/18 | 8:30-12:35 | 1.34-1.33 | HD 217196 | G5V | 0.69 | 1.62 |
| 259 Aletheia | 11/13/16 | 9:06-9:20 | 1.14-1.12 | HD285597 | G0 | 0.68 | 1.61 |
| 762 Pulcova | 09/9/20 | 8:41-11:00 | 1.17-1.01 | SAO_74411 | G5 | 0.67 | 1.58 |
| 165 Loreley | 08/11/20 | 11:50-14:45 | 1.17-1.88 | HD 207094 | G2V | 0.63 | 1.41 |
| 52 Europa | 06/22/17 | 5:51-7:19 | 1.23-1.13 | HD 133584 | G1V | -- | -- |
| 423 Diotima | 12/16/15 | 12:43-13:09 | 1.01-1.04 | SAO 79579 | G5 | 0.64 | 1.46 |
| 241 Germania | 02/04/20 | 13:20-16:10 | 1.13-2.39 | SAO 118200 | G5 | 0.71 | 1.86 |
| 360 Carlova | 01/04/17 | 5:32-7:38 | 1.52-1.06 | HD 244686 | G0 | 0.66 | 1.55 |

[*]B–V and V–K represent stars' colors.

**Table 5**. Observing parameters for asteroids observed in this study with the Prism mode of SpeX at NASA IRTF. The prism spectrum of asteroid 762 Pulcova is from (*10*).

| Asteroid | Date (UT) | Time (UT) | Airmass | Standard star | Spectral type | B-V | V-K |
|---|---|---|---|---|---|---|---|
| 94 Aurora | 12/29/20 | 14:04-14:50 | 1.00-1.03 | HD 86043 | G5 | 0.60 | 1.41 |
| 372 Palma | 08/21/16 | 13:33-13:40 | 1.59-1.60 | HD 210105 | G2V | 0.62 | 1.50 |
| 431 Nephele | 12/29/20 | 10:03-12:07 | 1.10-1.29 | SAO 98993 | G0 | 0.63 | 1.59 |
| 259 Aletheia | 11/13/16 | 7:56-8:10 | 1.40-1.43 | HD 285597 | G0 | 0.68 | 1.61 |
| 165 Loreley | 12/29/20 | 4.53-5.37 | 1.46-1.38 | HD 211567 | G5V | 0.61 | 1.46 |
| 52 Europa | 12/29/20 | 6.25-7.34 | 1.82-1.60 | HD 258906 | G5 | 0.66 | 1.54 |
| 423 Diotima | 12/29/20 | 5:41-6.31 | 1.38-1.83 | HD 283933 | G0 | 0.62 | 1.62 |
| 241 Germania | 12/29/20 | 14:54-15:54 | 1.94-1.51 | HD 123715 | G3V | 0.64 | 1.52 |
| 360 Carlova | 01/04/17 | 8:40-8:48 | 1.01-1.01 | HD244686 | G0 | 0.66 | 1.55 |

*B–V and *V–K represent stars' colors.

**Table 6**: Parameters used for the calculation of radiogenic energy. Heat Sources: The element mass fractions are referenced in Eq 1. The element mass fractions refer to stable isotopes, the initial ratios are between unstable and stable isotopes of an element, and the decay energies are per particle. The number of atoms of the stable isotope per 1 kg of the primordial material is $f=xN_A/m_a$ with the relative mass fraction $x$ of the stable isotope, the molar mass of the radioactive isotope $m_a$ in kg, and the Avogadro number $N_A$.

| Isotope | $^{26}$Al | $^{60}$Fe | $^{40}$K | $^{232}$Th | $^{235}$U | $^{238}$U |
|---|---|---|---|---|---|---|
| Element mass fraction $x$ | $8.68 \cdot 10^{-3}$ | $1.81 \cdot 10^{-1}$ | $5.32 \cdot 10^{-4}$ | $2.96 \cdot 10^{-8}$ | $8.08 \cdot 10^{-9}$ | $8.08 \cdot 10^{-9}$ |
| Half-life $\lambda$ [years] | $7.17 \cdot 10^5$ | $2.62 \cdot 10^6$ | $1.25 \cdot 10^9$ | $1.41 \cdot 10^{10}$ | $7.04 \cdot 10^8$ | $4.47 \cdot 10^9$ |
| Initial ratio $Z$ | $5 \cdot 10^{-5}$ | $1.15 \cdot 10^{-8}$ | $1.50 \cdot 10^{-3}$ | 1.0 | $8 \cdot 10^{-3}$ | $9.92 \cdot 10^{-1}$ |
| Decay energy $E$ [J] | $4.99 \cdot 10^{-13}$ | $4.34 \cdot 10^{-13}$ | $1.11 \cdot 10^{-13}$ | $6.47 \cdot 10^{-12}$ | $7.11 \cdot 10^{-13}$ | $7.61 \cdot 10^{-12}$ |